\documentstyle[12pt,epsf]{article}

\newtheorem{remark}{\sl Remark}

\def\aligned#1{\begin{array}{#1}}
\def\endaligned{\end{array}}

\def\matrix#1{\begin{array}{#1}}
\def\endmatrix{\end{array}}
\def\pmatrix#1{\left(\begin{array}{#1}}
\def\endpmatrix{\end{array}\right)}

\def\D{\displaystyle}
\def\tr{\mathop{\hbox{\,\rm tr}}\nolimits}
\def\re{\mathop{\hbox{\,\rm Re}}\nolimits}
\def\im{\mathop{\hbox{\,\rm Im}}\nolimits}
\def\hc{\hbox{\bf C}}
\def\hr{\hbox{\bf R}}
\def\diag{\mathop{\hbox{\,\rm diag}}\nolimits}

\newif\ifinexp \inexpfalse
\def\I{\ifinexp \hbox{\hskip0.8pt\eightrm i} \else \hbox{\hskip1pt\rm i} \fi}

\newcount\egcount \egcount=1
\def\example#1{\vskip4pt
   \noindent{\bf Example \the\egcount: }#1\par\advance\egcount by 1}
\def\endexample{\vskip4pt}

\def\adS{2+1 dimensional anti-de Sitter space-time}
\def\YMH{Yang-Mills-Higgs}
\def\YMHB{Yang-Mills-Higgs-Bogomolny}

\def\BMPtype{BMP}

\def\figuretype{EPS}

\begin{document}

\title{Relation between the solitons of Yang-Mills-Higgs fields
in 2+1 dimensional Minkowski space-time and anti-de Sitter space-time}

\author{\small Zixiang Zhou\\
\small Institute of Mathematics, Fudan University,
Shanghai 200433, China\\
\small Email: zxzhou@guomai.sh.cn}
\date{}

\maketitle

\begin{abstract}
The \YMHB{} equations in both 2+1 dimensional Minkowski space-time
and \adS{} are known to be integrable and their soliton solutions
have already been obtained. In this paper we show that there is a
natural relation between the Lax pairs and soliton solutions in
these two space-times  when the curvature changes from $0$ to
$-1$. The change of the asymptotic behaviors of the solitons are
also discussed.
\end{abstract}

\section{Introduction}\label{sec1}

The \YMHB{} equations in both 2+1 dimensional Min\-kow\-ski
space-time and \adS{} are known to be integrable
\cite{bib:Atiyah,bib:Hitchinbook,bib:Ward,bib:Wardnew}. There are
several ways to solve them explicitly. Darboux transformation
method is one of them, which gives an easy way to obtain explicit
soliton solutions \cite{bib:GuYMH,bib:ZhouYMH}. Since the Lax
pairs in both Minkowski and anti-de Sitter cases are known, the
Darboux transformations can be constructed separately in these
two cases.

The standard anti-de Sitter space-time has curvature
$-1$. Naturally we can consider the anti-de Sitter space-time
with constant curvature $-1/\rho^2$ $(\rho>0)$. When
$\rho\to+\infty$, the space-time tends to the Minkowski space-time.
In this paper, we consider the following problem: When $\rho$
changes from $1$ to $+\infty$, whether the solitons in the
anti-de Sitter space-time change to solitons in the Minkowski
space-time? 

In Section~\ref{sec2}, the \YMHB{} equations and their Lax pairs
for general $\rho$ are considered. When $\rho=1$ and
$\rho\to+\infty$, they become the known equations and their Lax
pairs for Minkowski and anti-de Sitter cases. In
Section~\ref{sec3}, the Darboux transformation is discussed.
Using the Darboux transformation, we construct solitons in
$SU(2)$ case in Section~\ref{sec4} and give some examples. When
$\rho$ changes from $1$ to $+\infty$, the shape of solitons
changes a lot. However, when the coordinates of the space-time
depend on $\rho$ suitably, the position of the solitons keeps in a
finite region and the solitons in part of the anti-de Sitter
space-time change to the solitons in the Minkowski space-time. 

\section{Yang-Mills-Higgs-Bogomolny equations\\ and their Lax
pairs}\label{sec2}
Let $M$ be a three dimensional Lorentz manifold with metric $g=(g_{\mu\nu})$.
$\{A_\mu\,|\,\mu=1,2,3\}$ is a gauge potential and $\Phi$ is a
(scalar) Higgs field, both of which are valued in the Lie algebra
of an $N\times N$ matrix Lie group $G$.

The \YMHB{} equation \cite{bib:Atiyah,bib:Hitchin1} is
\begin{equation}
   D\Phi=*F,
   \label{eq:YMeq}
\end{equation}
or, written in terms of the components,
\begin{equation}
   D_\mu\Phi=\frac 1{2\sqrt{|g|}}g_{\mu\nu}
   \epsilon^{\nu\alpha\beta}F_{\alpha\beta}
   \label{eq:YMeqcompo}
\end{equation}
where the action of the covariant derivative
$D_\mu=\partial_\mu+A_\mu$ on $\Phi$ is
$D_\mu\Phi=\partial_\mu\Phi+[A_\mu,\Phi]$,
$\partial_\mu=\partial/\partial x^\mu$. $\{F_{\mu\nu}\}$ is the
curvature corresponding to $\{A_\mu\}$, hence $F_{\mu\nu}=[D_\mu,D_\nu]$.

The \adS{} of constant curvature $-1/\rho^2$ $(\rho>0)$ is
the hyperboloid
$U^2+V^2-X^2-Y^2=\rho^2$
in $\hr^{2,2}$ with the metric
\begin{equation}
   ds^2=-dU^2-dV^2+dX^2+dY^2.
\end{equation}
By defining
\begin{equation}
   r=\frac \rho{U+X}-\rho+1,\quad
   x=\frac Y{U+X},\quad
   t=-\frac V{U+X},
   \label{eq:xrt}
\end{equation}
a part of the \adS{} with $U+X>0$ is represented by the
Poincar\'e coordinates $(r,x,t)$ with $r>-\rho+1$ and the metric is
\begin{equation}
   ds^2=\frac{\rho^2}{(r+\rho-1)^2}(-dt^2+dr^2+dx^2)
    =\frac{\rho^2}{(r+\rho-1)^2}(dr^2+du\,dv)
   \label{eq:metric}
\end{equation}
where $u=x+t$, $v=x-t$.
Clearly, when $\rho\to+\infty$, the metric on this part of the \adS{}
tends to the flat Minkowski metric on the whole $\hr^{2,1}$. In
order to consider the change of the solitons with respect to $\rho$, we only
need to consider the solutions in this part.

With the metric (\ref{eq:metric}) and the orientation $(v,u,r)$,
(\ref{eq:YMeqcompo}) becomes
\begin{equation}
   D_u\Phi=\frac{r+\rho-1}{\rho}F_{ur},\quad
   D_v\Phi=-\frac{r+\rho-1}{\rho}F_{vr},\quad
   D_r\Phi=-\frac{2(r+\rho-1)}{\rho}F_{uv}.
   \label{eq:EQ}
\end{equation}

When $\rho=1$, it is reduced to
\begin{equation}
   D_u\Phi=rF_{ur},\quad
   D_v\Phi=-rF_{vr},\quad
   D_r\Phi=-2rF_{uv}.
   \label{eq:EQadS}
\end{equation}
\cite{bib:Wardnew} showed that it had a Lax pair
\begin{equation}
   (rD_r+\Phi-2(\zeta-u)D_u)\psi=0,\qquad
   \D\left(2D_v+\frac{\zeta-u}rD_r-\frac{\zeta-u}{r^2}\Phi\right)\psi=0
   \label{eq:LPadS}
\end{equation}
where $D_\mu\psi=\partial_\mu\psi+A_\mu\psi$ and $\zeta$ is a
complex spectral parameter. That is, (\ref{eq:EQadS}) is the
integrability condition of the over-determined system
(\ref{eq:LPadS}).

When $\rho>0$, the \YMHB{} equation (\ref{eq:EQ}) can be derived
from (\ref{eq:EQadS}) by substituting $r\to r+\rho-1$ and
$\Phi\to\rho\Phi$. Moreover, since $\zeta$ is a constant in
(\ref{eq:LPadS}), we can replace $\zeta$ by $\rho\zeta$.
After the substitution
\begin{equation}
   r\to r+\rho-1,\quad
   \Phi\to\rho\Phi,\quad
   \zeta\to\rho\zeta,
   \label{eq:subst}
\end{equation}
(\ref{eq:LPadS}) leads to the Lax pair of (\ref{eq:EQ}):
\begin{equation}
  \aligned{l}
   ((r+\rho-1)D_r+\rho\Phi-2(\rho\zeta-u)D_u)\psi=0,\\
   \D\left(2D_v+\frac{\rho\zeta-u}{r+\rho-1}D_r
    -\frac{\rho(\rho\zeta-u)}{(r+\rho-1)^2}\Phi\right)\psi=0.
   \endaligned \label{eq:LP}
\end{equation}
It is easy to check directly that the integrability condition of
(\ref{eq:LP}) is the \YMHB{} equation (\ref{eq:EQ}).

When $\rho\to+\infty$, the metric (\ref{eq:metric}) becomes the
standard Minkowski metric
\begin{equation}
   ds^2=-dt^2+dr^2+dx^2=dr^2+du\,dv,
   \label{eq:metricMin}
\end{equation}
the \YMHB{} equation (\ref{eq:EQ}) becomes
\begin{equation}
   D_u\Phi=F_{ur},\quad
   D_v\Phi=-F_{vr},\quad
   D_r\Phi=-2F_{uv},
   \label{eq:EQMin}
\end{equation}
and the Lax pair (\ref{eq:LP}) becomes
\begin{equation}
   \aligned{l}
   (D_r+\Phi-2\zeta D_u)\psi=0,\\
   (2D_v+\zeta D_r-\zeta\Phi)\psi=0.
   \label{eq:LPMin}
   \endaligned
\end{equation}

\begin{remark}
If we substitute
\begin{equation}
   r\to x,\quad \D\zeta\to\frac 1\lambda,\quad
   u\to y+t,\quad v\to y-t,
   \label{eq:substMin}
\end{equation}
then (\ref{eq:LPMin}) is changed to
\begin{equation}
   (\lambda D_x-D_t-D_y+\lambda\Phi)\psi=0,\qquad
   (\lambda D_t-\lambda D_y-D_x+\Phi)\psi=0,
\end{equation}
which is just the Lax pair given by \cite{bib:Hitchinbook}.
\end{remark}

\section{Darboux transformations}\label{sec3}

For $\rho\to+\infty$ and $\rho=1$, \cite{bib:GuYMH} and
\cite{bib:ZhouYMH} gave the construction of the Darboux matrix
separately based on a general method \cite{bib:GuDT}. Here we
show that these are the two special cases for general $\rho$.

For $\rho=1$, the Darboux transformation is given as follows
\cite{bib:ZhouYMH}. Let $Z=\diag(\zeta_1,\cdots,\zeta_N)$ be a
diagonal matrix and satisfies
\begin{equation}
   \D\partial_rZ-\frac{2(Z-u)}{r}(\partial_uZ)=0,\qquad
   \D\partial_vZ+\frac{Z-u}{2r}(\partial_rZ)=0,
   \label{eq:ZeqadS}
   \end{equation}
$H=(h_1,\cdots,h_N)$ where $h_j$ is a solution of
(\ref{eq:LPadS}) with $\zeta=\zeta_j$, then $G=\zeta-HZH^{-1}$ is
a Darboux matrix for (\ref{eq:LPadS}). That is, for any solution
$\psi$ of the Lax pair (\ref{eq:LPadS}), $\widetilde\psi=G\psi$
satisfies
\begin{equation}
   (r\widetilde D_r+\widetilde\Phi-2(\zeta-u)\widetilde D_u)
    \widetilde\psi=0,\qquad
   \D\left(2\widetilde D_v+\frac{\zeta-u}r\widetilde D_r
    -\frac{\zeta-u}{r^2}\widetilde\Phi\right)\widetilde\psi=0
\end{equation}
where $\widetilde D_\mu=\partial_\mu+\widetilde A_\mu$ and
$\widetilde\Phi,\widetilde A_\mu$ are other functions in the Lie
algebra of $G$.

When $\rho>1$, similar conclusion is obtained by the substitution
(\ref{eq:subst}) and $Z\to\rho Z$. Hence the Darboux matrix is
given by
\begin{equation}
   G(r,u,v,\zeta)=\zeta-\frac u\rho-S(r,u,v),\qquad
   S(r,u,v)=H\left(Z-\frac u\rho\right)H^{-1}
\end{equation}
where $Z=\diag(\zeta_1,\cdots,\zeta_N)$
satisfies
\begin{equation}
   \D\partial_rZ-\frac{2(\rho Z-u)}{r+\rho-1}\partial_uZ=0,\qquad
   \D\partial_vZ+\frac{\rho Z-u}{2(r+\rho-1)}\partial_rZ=0,
   \label{eq:Zeq}
   \end{equation}
$H=(h_1,\cdots,h_N)$ and $h_j$ is a solution of (\ref{eq:LP})
with $\zeta=\zeta_j$.
It can be checked that $S$ satisfies
\begin{equation}
   \begin{array}{l}
   (r+\rho-1)(\partial_rS+[A_r,S])-2\rho(\partial_uS+[A_u,S])S
    +\rho[\Phi,S]-2S=0,\\
   \D 2(\partial_vS+[A_v,S])+\frac\rho{r+\rho-1}(\partial_rS+[A_r,S])S
    -\frac{\rho^2}{(r+\rho-1)^2}[\Phi,S]S=0.
   \end{array}
\end{equation}

By direct computation, we know that for any
solution $\psi$ of (\ref{eq:LP}), $\widetilde\psi=G\psi$ satisfies
\begin{equation}
  \aligned{l}
   ((r+\rho-1)\widetilde D_r+\rho\widetilde\Phi
    -2(\rho\zeta-u)\widetilde D_u)\widetilde\psi=0,\\
   \D\left(2\widetilde D_v+\frac{\rho\zeta-u}{r+\rho-1}\widetilde D_r
    -\frac{\rho(\rho\zeta-u)}{(r+\rho-1)^2}\widetilde\Phi\right)
    \widetilde\psi=0
   \endaligned
\end{equation}
with $\widetilde D_\mu=\partial_\mu+\widetilde A_\mu$ $(\mu=u,v,r)$,
\begin{equation}
   \begin{array}{l}
   \widetilde A_u=A_u,\\
   \D\widetilde A_v=A_v+\frac{\rho}{2(r+\rho-1)}(\partial_rS+[A_r,S])
    -\frac{\rho^2}{2(r+\rho-1)^2}[\Phi,S],\\
   \D\widetilde A_r=A_r-\frac{1+\rho(\partial_uS+[A_u,S])}{r+\rho-1},\\
   \D\widetilde\Phi=\Phi-\frac{1+\rho(\partial_uS+[A_u,S])}{\rho}.
   \end{array}\label{eq:DTA}
\end{equation}
Hence $G$ is really a Darboux matrix for (\ref{eq:LP}).


According to (\ref{eq:Zeq}), each $\zeta_j$ $(j=1,\cdots,N)$ is a
constant or a non-constant solution of
\begin{equation}
   \D\partial_r\zeta-\frac{2(\rho \zeta-u)}{r+\rho-1}\partial_u\zeta=0,
   \qquad
   \D\partial_v\zeta+\frac{\rho \zeta-u}{2(r+\rho-1)}\partial_r\zeta=0.
   \label{eq:zeta}
   \end{equation}
The general non-constant solution is given implicitly by
\begin{equation}
   v-\frac{(r+\rho-1)^2}{\rho\zeta-u}=C_1(\zeta,\rho)
   \label{eq:nonconst1}
\end{equation}
where $C_1$ is an arbitrary function, which is meromorphic to
$\zeta$ and smooth to $\rho\in(0,+\infty)$.

In order to consider the limit for $\rho\to+\infty$, we rewrite
(\ref{eq:nonconst1}) as
\begin{equation}
   v-\frac{(r+\rho-1)^2}{\rho\zeta-u}+\frac{\rho-1}{\zeta}=C(\zeta,\rho).
   \label{eq:nonconst}
\end{equation}
Here $C(\zeta,\rho)$ is also an arbitrary function, which is
holomorphic to $\zeta$ and smooth to $\rho$. Moreover, suppose that
$\D\lim_{\rho\to+\infty}C(\zeta,\rho)$ exists.

When $\rho=1$, (\ref{eq:nonconst}) becomes
\begin{equation}
   v-\frac{r^2}{\zeta-u}=C(\zeta,1)
\end{equation}
which is given by \cite{bib:ZhouYMH}. When $\rho\to+\infty$,
(\ref{eq:nonconst}) becomes
\begin{equation}
  v-\frac{u}{\zeta^2}-\frac{2r}{\zeta}=C(\zeta,+\infty)-\frac 1\zeta.
\end{equation}

When the group $G=U(N)$, there should be more constraints on
$\zeta_j$'s and $h_j$'s in the construction of Darboux matrix.
They are:
\begin{equation}
   \begin{array}{l}
   \zeta_j=\zeta_0\hbox{ or }\bar\zeta_0\hbox{ for certain fixed }\zeta_0,\\
   h_j^*h_k=0\hbox{ if }\zeta_j\ne \zeta_k,
   \end{array}
   \label{eq:uNconsition}
\end{equation}
as mentioned in \cite{bib:GuYMH,bib:ZhouYMH}. If so, after the
Darboux transformation, $\widetilde\Phi\in u(N)$, $\widetilde
A_\mu\in u(N)$ provided that $\Phi\in u(N)$,
$A_\mu\in u(N)$.

\section{Soliton solutions in SU(2) case}\label{sec4}

Single soliton solutions are given by Darboux transformations from
the trivial seed solution $\Phi=0$, $A_u=A_v=A_r=0$. In the
construction of $S=H(Z-u/\rho)H^{-1}$,
$Z=\diag(\zeta_1,\cdots,\zeta_N)$ where $\zeta_j$ is a constant
or a non-constant solution of (\ref{eq:zeta}), $h_j$ is a column
solution of (\ref{eq:LP}) with $\zeta=\zeta_j$.

With the action of the Darboux matrix $G=\zeta-u/\rho-S$,
(\ref{eq:DTA}) gives
\begin{equation}
   \D\widetilde A_u=0,\qquad
    \widetilde A_v=\frac{\rho\partial_rS}{2(r+\rho-1)},\qquad
   \D\widetilde A_r=-\frac{1+\rho\partial_uS}{r+\rho-1},\qquad
    \widetilde\Phi=-\frac{1+\rho\partial_uS}{\rho}.
\end{equation}

Here we only consider the case where all $\zeta_j$'s are
constants. When $\zeta_j$'s are non-constant solutions of
(\ref{eq:zeta}), we can obtain solutions in similar ways.
However, in the latter case, solutions may only defined when $t$
is large then some constant \cite{bib:ZhouYMH}.
Now $h_j$ satisfies
\begin{equation}
   \D\partial_rh_j-\frac{2(\rho\zeta_j-u)}{r+\rho-1}\partial_uh_j=0,\qquad
   \D\partial_vh_j+\frac{\rho\zeta_j-u}{2(r+\rho-1)}\partial_rh_j=0.
\end{equation}
The general solution is
\begin{equation}
   h_j=\omega(\zeta_j)
\end{equation}
where
\begin{equation}
   \omega(\zeta)=v-\frac{(r+\rho-1)^2}{\rho\zeta-u}+\frac{\rho-1}{\zeta}.
\end{equation}
When $\rho=1$,
\begin{equation}
   \omega(\zeta)=v-\frac{r^2}{\zeta-u},
   \label{eq:omega1}
\end{equation}
which is the same as the result in \cite{bib:ZhouYMH}.
When $\rho\to+\infty$,
\begin{equation}
   \omega(\zeta)\to v-\frac{u}{\zeta^2}-\frac{2r}{\zeta}+\frac{1}{\zeta}.
   \label{eq:omegainfty}
\end{equation}
With the substitution (\ref{eq:substMin}),
\begin{equation}
   \omega(\lambda^{-1})\to (1-\lambda^2)y-(1+\lambda^2)t-2\lambda x+\lambda.
\end{equation}
This coincides with \cite{bib:GuYMH}.

When $G=SU(2)$, the conditions (\ref{eq:uNconsition}) should be satisfied.
Hence we want $\zeta_1=\zeta_0$, $\zeta_2=\bar\zeta_0$
for some
$\zeta_0\in\hc$ and
\begin{equation}
   H=\pmatrix{cc}\alpha(\tau) &-\overline{\beta(\tau)}\\
   \beta(\tau) &\overline{\alpha(\tau)}\endpmatrix
\end{equation}
where $\alpha$, $\beta$ are two holomorphic functions of
$\tau=\omega(\zeta_0)$. Let
$\sigma(\tau)=\beta(\tau)/\alpha(\tau)$, then
\begin{equation}
   S=\frac{\zeta_0-\bar\zeta_0}{1+|\sigma|^2}
   \pmatrix{cc} 1 &\bar\sigma\\ \sigma &|\sigma|^2\endpmatrix+\bar\zeta_0
   -\frac u\rho.
   \label{eq:egS}
\end{equation}
\begin{equation}
   \widetilde\Phi=-\partial_uS-\frac 1\rho
    =\frac{\zeta_0-\bar\zeta_0}{(1+|\sigma|^2)^2}
   \pmatrix{cc} (|\sigma|^2)_u &\bar\sigma^2\sigma_u-\bar\sigma_u\\
    \sigma^2\bar\sigma_u-\sigma_u &-(|\sigma|^2)_u\endpmatrix
   \label{eq:egPhi}
\end{equation}
and
\begin{equation}
   -\tr\widetilde\Phi^2=\frac{8(\im \zeta_0)^2}{(1+|\sigma|^2)^2}|
    \partial_u\sigma|^2.
   \label{eq:Energy}
\end{equation}

When $\sigma(z)$ is a given meromorphic function of $z$ which is
independent of $\rho$, then by (\ref{eq:omegainfty}) and (\ref{eq:omega1}),
\begin{equation}
   \sigma(\tau)|_{\rho\to+\infty}=\sigma\left(v-\frac u{\zeta_0^2}
    -\frac{2r}{\zeta_0}+\frac 1{\zeta_0}\right),
   \quad
   \sigma(\tau)|_{\rho=1}=\sigma\left(v-\frac{r^2}{\zeta_0-u}\right).
   \label{eq:sigmalimit}
\end{equation}
Hence when $\rho\to+\infty$ and $\rho=1$, the solutions tend to the
soliton solutions in the Minkowski and anti-de Sitter
space-time respectively.

These are single soliton solutions. Each solution depends on a
complex constant $\zeta_0$ and a meromorphic function $\sigma$.
Multi-soliton solutions can be constructed by successive Darboux
transformations \cite{bib:GuYMH,bib:ZhouYMH}. For simplicity,
here we only consider the change of single soliton solutions with
respect to $\rho$.

\example{$\sigma(\tau)$ is a polynomial of $\tau$ without
multiple zero}

In this case, the solutions are always localized. When $\rho=1$,
the behavior of the asymptotic solution as $t\to\infty$ varies
according to the roots of $\sigma(\tau)$ \cite{bib:ZhouYMH}.
Suppose $\tau_0$ is a  root of $\sigma(\tau)$, then (1) if
$|\im\tau_0|<<1$, it corresponds to a ridge in the graph of
$-\tr\widetilde\Phi^2$; (2) if $\im\tau_0>>1$, it corresponds to
a peak; (3) if $\im\tau_0<<-1$, it corresponds to nothing.
However, when $\rho\to+\infty$, each root of $\sigma(\tau)$
corresponds to a peak \cite{bib:GuYMH}. The following figures
1--5 show the change of the solution with respect to $\rho$ for
fixed $t=10$, where $\zeta_0=2\I$,
\begin{equation}
   \sigma(\tau)=(\tau-2)(\tau-6)(\tau+6)(\tau-2\I)(\tau-6\I)(\tau+6\I).
\end{equation}
In these figures the vertical axis is $(-\tr\widetilde\Phi^2)^{1/16}$.
\endexample

\unitlength=1mm

\vbox{%
\ifx\figuretype\BMPtype
{
\begin{picture}(60,50)
\put(6,50){\special{em:graph poly1.bmp}}
\end{picture}
\vskip0.6cm
}
\else
{
\vskip-1cm
\epsffile[-60 0 400 200]{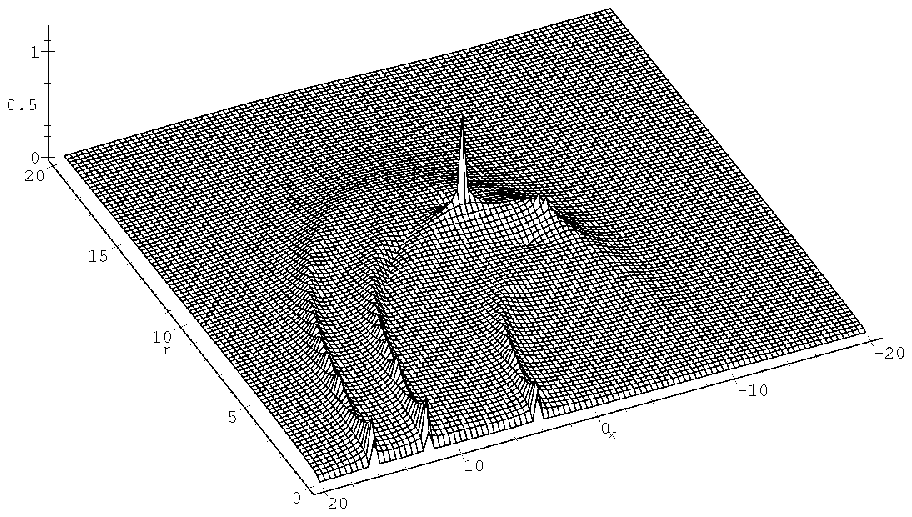}
}
\fi
\gdef\captionone{$\rho=1$}
\hbox to \hsize{\hfill\scriptsize Fig.~\ref{fig:poly1}.
\captionone \hfill\hfill}}

\vbox{%
\ifx\figuretype\BMPtype
{
\begin{picture}(60,50)
\put(6,50){\special{em:graph poly2.bmp}}
\end{picture}
\vskip0.6cm
}
\else
{
\vskip-1cm
\epsffile[-60 0 400 200]{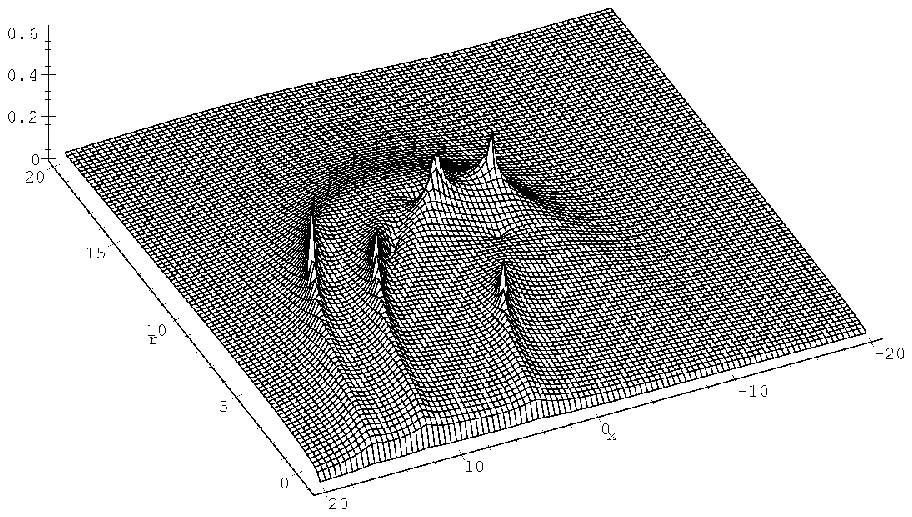}
}
\fi
\gdef\captiontwo{$\rho=2$}
\hbox to \hsize{\hfill\scriptsize Fig.~\ref{fig:poly2}.
\captiontwo\hfill\hfill}} 

\vbox{%
\ifx\figuretype\BMPtype
{
\begin{picture}(60,50)
\put(6,50){\special{em:graph poly3.bmp}}
\end{picture}
\vskip0.6cm
}
\else
{
\vskip-1cm
\epsffile[-60 0 400 200]{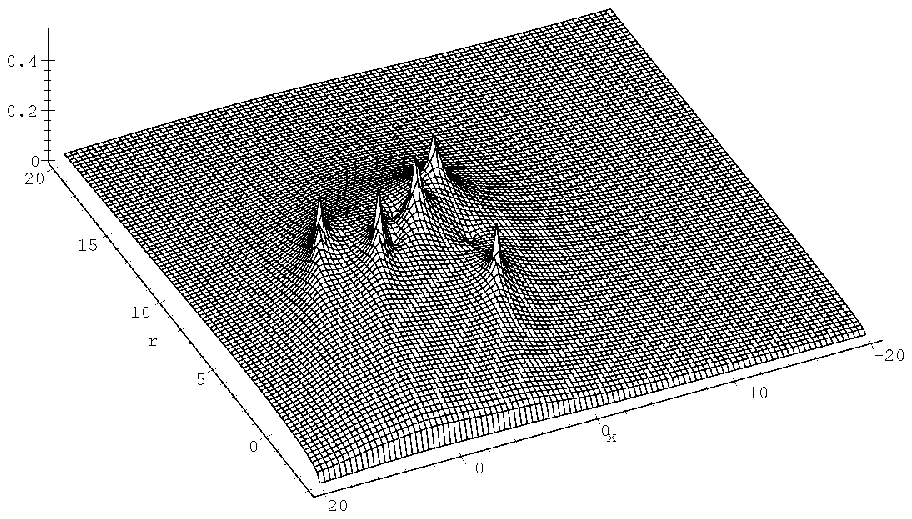}
}
\fi
\gdef\captionthree{$\rho=5$}
\hbox to \hsize{\hfill\scriptsize Fig.~\ref{fig:poly3}.
\captionthree \hfill\hfill}}

\vbox{%
\ifx\figuretype\BMPtype
{
\begin{picture}(60,50)
\put(6,50){\special{em:graph poly4.bmp}}
\end{picture}
\vskip0.6cm
}
\else
{
\vskip-1cm
\epsffile[-60 0 400 200]{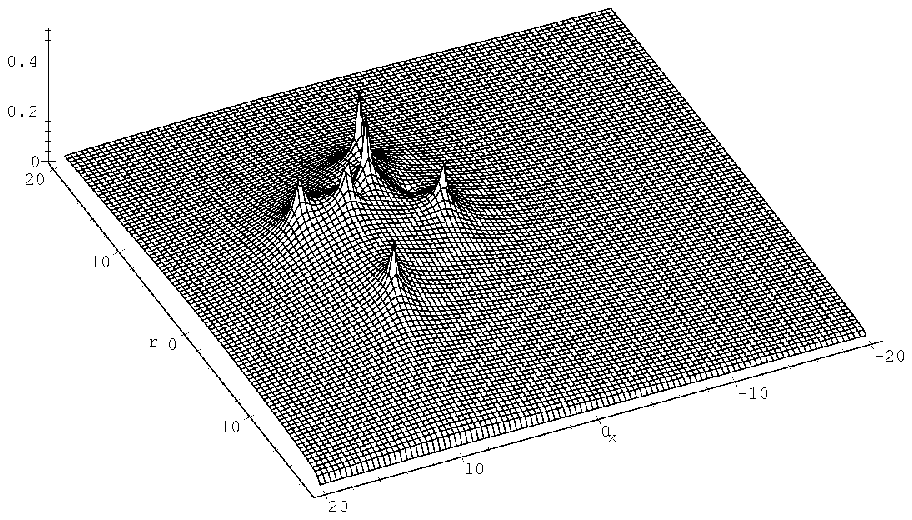}
}
\fi
\gdef\captionfour{$\rho=20$}
\hbox to \hsize{\hfill\scriptsize Fig.~\ref{fig:poly4}.
\captionfour\hfill\hfill}} 

\vbox{%
\ifx\figuretype\BMPtype
{
\begin{picture}(60,50)
\put(6,50){\special{em:graph poly5.bmp}}
\end{picture}
\vskip0.6cm
}
\else
{
\vskip-1cm
\epsffile[-60 0 400 200]{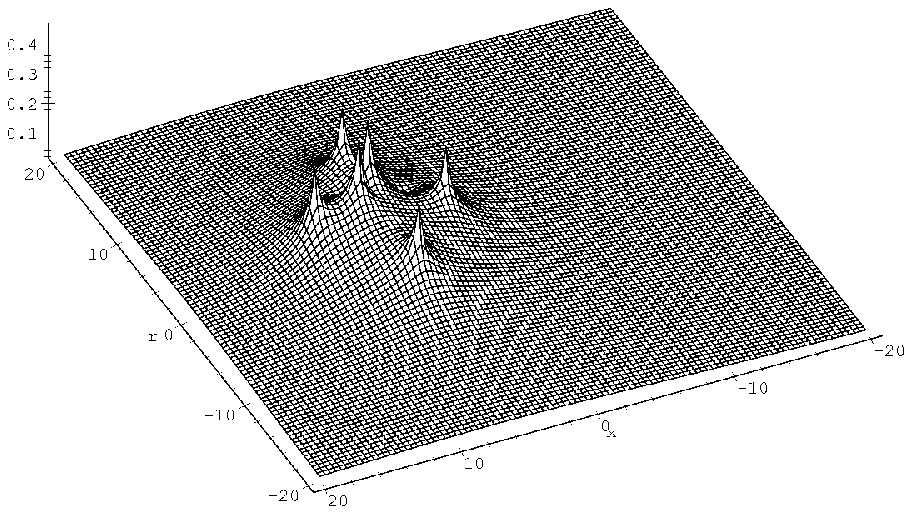}
}
\fi
\gdef\captionfive{$\rho=+\infty$}
\hbox to \hsize{\hfill\scriptsize Fig.~\ref{fig:poly5}.
\captionfive \hfill\hfill}}

\example{$\sigma(\tau)=\sin(\tau/20)$}
For both finite and infinite $\rho$, the solution is always
non-localized. For finite $\rho$, it behaves very complicated.
However, for infinite $\rho$, (\ref{eq:sigmalimit}) shows that
the solution is invariant if $(x,r)$ is changed to $(x',r')$ with
$\re[(1-\zeta_0^{-2})(x'-x)-2\zeta_0^{-1}(r'-r)]=40\pi k$ ($k$ is
an arbitrary integer). Hence the solution is periodic in one
direction. Figures 6--8 show this solution for
$\rho=1,30,+\infty$ with $t=10$, $\zeta_0=2\I$.
In these figures the vertical axis is $(-\tr\widetilde\Phi^2)^{1/8}$.
\endexample

\vskip0.5cm
\vbox{%
\ifx\figuretype\BMPtype
{
\begin{picture}(60,50)
\put(6,50){\special{em:graph sin1.bmp}}
\end{picture}
\vskip1.3cm
}
\else
{
\vskip-1cm
\epsffile[-60 0 400 200]{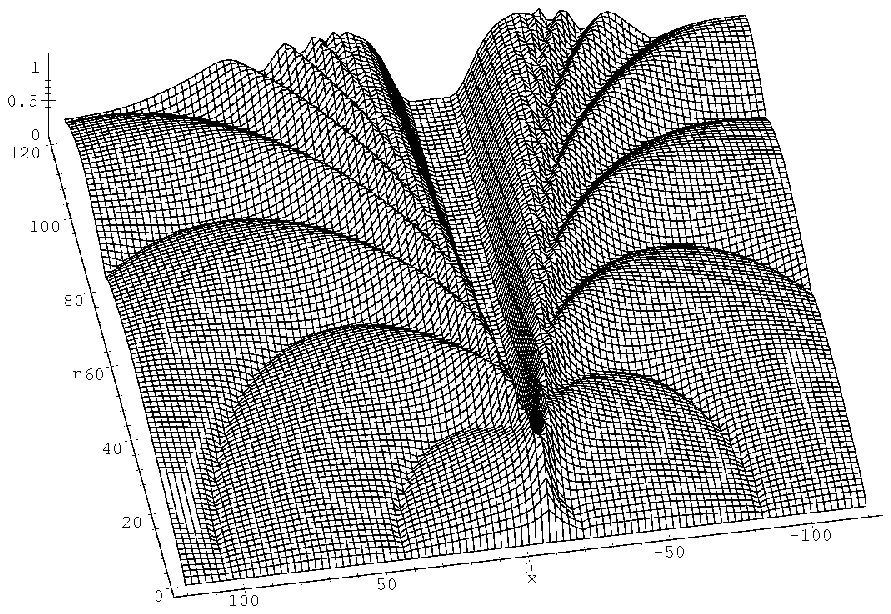}
\vskip-0.4cm
}
\fi
\gdef\captionsix{$\rho=1$}
\hbox to \hsize{\hfill\scriptsize Fig.~\ref{fig:sin1}.
\captionsix \hfill\hfill}}

\vbox{%
\ifx\figuretype\BMPtype
{
\begin{picture}(60,50)
\put(6,50){\special{em:graph sin2.bmp}}
\end{picture}
\vskip1.3cm
}
\else
{
\vskip-0.3cm
\epsffile[-60 0 400 200]{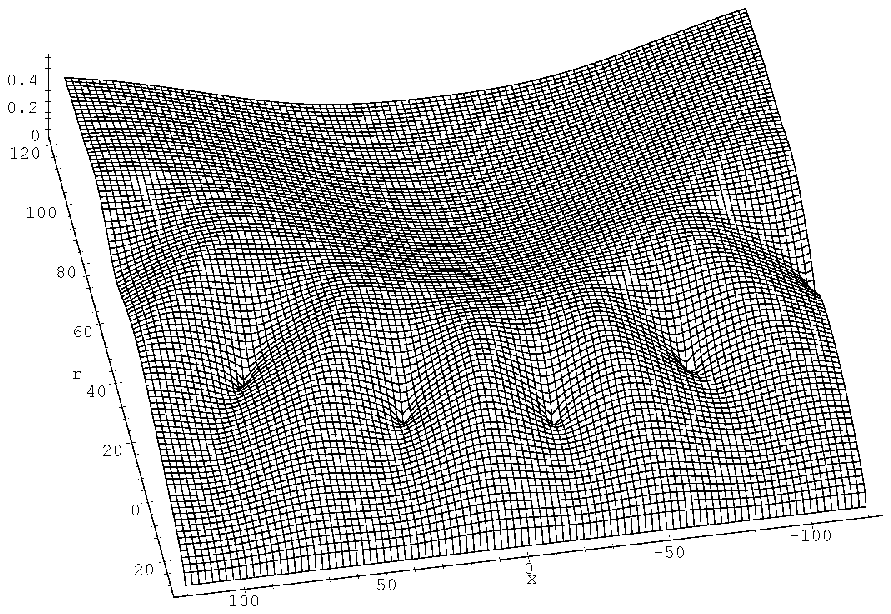}
\vskip-0.4cm
}
\fi
\gdef\captionseven{$\rho=30$}
\hbox to \hsize{\hfill\scriptsize Fig.~\ref{fig:sin2}.
\captionseven \hfill\hfill}}

\vbox{%
\ifx\figuretype\BMPtype
{
\begin{picture}(60,50)
\put(6,50){\special{em:graph sin3.bmp}}
\end{picture}
\vskip1.3cm
}
\else
{
\vskip-0.5cm
\epsffile[-60 0 400 200]{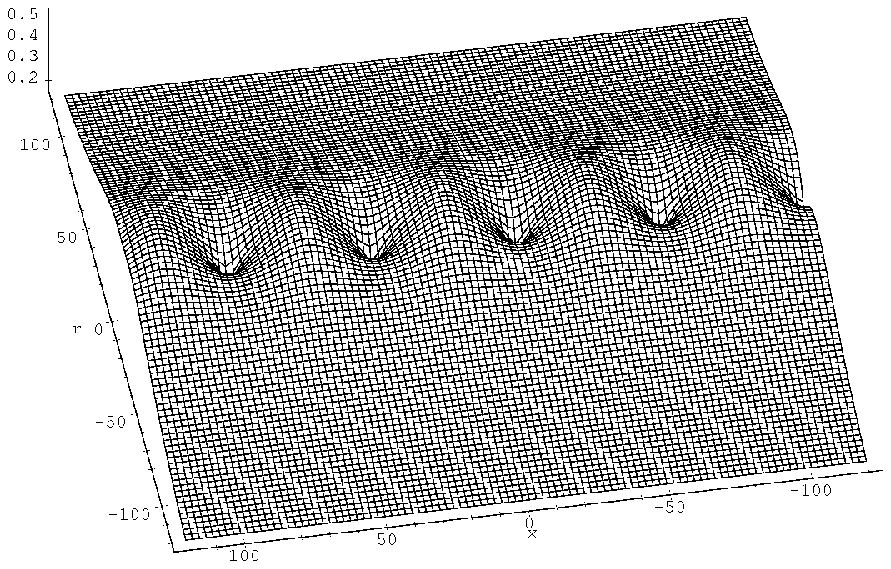}
\vskip-0.4cm
}
\fi
\gdef\captioneight{$\rho=+\infty$}
\hbox to \hsize{\hfill\scriptsize Fig.~\ref{fig:sin3}.
\captioneight \hfill\hfill}}

\section*{Acknowledgements}

This work was supported by the Special Funds for Chinese Major State Basic Research
Projects, the Doctoral Program Foundation, the
Trans-century Training Program Foundation for the Talents and the
Foundation for University Key Teacher by the Ministry of
Education of China.

The author is very grateful to Prof.~M.~F.~Atiyah for his
suggestion of studying the problem in this paper. He is also
grateful to Prof.~C.~H.~Gu and Prof.~M.~L.~Ge for their helpful
suggestions.


\newpage

\begin{figure}
\caption{\captionone}
\label{fig:poly1}
\end{figure}

\begin{figure}
\caption{\captiontwo}
\label{fig:poly2}
\end{figure}

\begin{figure}
\caption{\captionthree}
\label{fig:poly3}
\end{figure}

\begin{figure}
\caption{\captionfour}
\label{fig:poly4}
\end{figure}

\begin{figure}
\caption{\captionfive}
\label{fig:poly5}
\end{figure}

\begin{figure}
\caption{\captionsix}
\label{fig:sin1}
\end{figure}

\begin{figure}
\caption{\captionseven}
\label{fig:sin2}
\end{figure}

\begin{figure}
\caption{\captioneight}
\label{fig:sin3}
\end{figure}

\end{document}